\documentstyle[12pt]{article}
\newcommand{\prt}{\partial}

\begin{document}
\renewcommand{\theequation}{\thesection.\arabic{equation}}
\newcommand{\beq}{\begin{equation}}
\newcommand{\eeq}{\end{equation}}
\newcommand{\sibsi}{\sigma_{\mu}\bar{\sigma_{\nu}}}
\newcommand{\bsisi}{\bar{\sigma_{\mu}}\sigma_{\nu}}
\newcommand{\F}{F_{\mu\nu}}
\newcommand{\al}{\alpha}
\newcommand{\dal}{\dot{\alpha}}
\newcommand{\bt}{\beta}
\newcommand{\ep}{\epsilon}
\newcommand{\dl}{\delta}
\newcommand{\bep}{\bar{\epsilon}}
\newcommand{\bsi}{\bar{\sigma}}
\newcommand{\bbt}{\bar{\beta}}
\newcommand{\bla}{\bar{\lambda}}
\newcommand{\bth}{\bar{\theta}}
\newcommand{\bet}{\bar{\eta}}
\newcommand{\bPh}{\bar{\Phi}}
\newcommand{\bph}{\bar{\phi}}
\newcommand{\bPsi}{\bar{\Psi}}
\newcommand{\si}{\sigma}
\newcommand{\nbl}{\nabla}
\newcommand{\qnbl}{\nabla^{2}}
\newcommand{\bnbl}{\bar{\nabla}}
\newcommand{\qbnb}{\bar{\nabla}^{2}}
\newcommand{\tht}{\theta}
\newcommand{\qpi}{\pi^{2}}
\newcommand{\qrho}{\rho^{2}}
\newcommand{\qd}{d^{2}}
\newcommand{\td}{d^{3}}
\newcommand{\fd}{d^{4}}
\newcommand{\qg}{g^{2}}
\newcommand{\la}{\lambda}
\newcommand{\La}{\Lambda}
\newcommand{\dbt}{\dot{\beta}}
\newcommand{\nd}[1]{/\hspace{-0.5em} #1}

\title{Instanton-induced Effective Lagrangian in the Seiberg-Witten
Model}
\author{Alexei Yung \thanks{Permanent address: Petersburg
 Nuclear Physics Institute
Gatchina, St.Petersburg 188350, Russia, yung@thd.pnpi.spb.ru}\\
Physics Department, University College of Swansea \\
 Swansea SA2 8PP, UK, a.yung@swansea.ac.uk}
\date{May 1996}
\maketitle

\begin{abstract}
The N=2 supersymmetric gauge theory  with gauge group
SU(2) is considered. The  instanton field is calculated explicitly using
the superfield formalism. The instanton-induced effects are encoded
in the effective vertex in the Lagrangian. 
This vertex produces the large distance expansion of the
low energy effective Lagrangian in derivatives of fields.
The leading term in this expansion coincides with 
 the one-instanton-induced term 
of the Seiberg-Witten exact solution of the model. 
All orders of  corrections in  derivatives
of fields are also calculated.
\end{abstract}
\vspace*{2cm}
SWAT/ 96/111\\
hep-th 9605096
\newpage

\section{Introduction}

A lot of progress has been made last time in the study of supersymmetric
gauge theories in four dimensions. The use of electromagnetic
duality \cite{O} and holomorphy allows Seiberg and Witten to solve exactly
for the low energy effective Lagrangian of the N=2 SUSY SU(2) gauge
theory with \cite{SW2} and without matter hypermultiplets \cite{SW}.
The long standing problems of confinement and chiral symmetry breaking
get its natural explanation via condensation of dual fields \cite{SW,SW2}.
The extrapolation of these ideas to N=1 SUSY and non-abelian gauge groups
is under intensive study \cite{S94,S95,APS}.

The exact solution for the low energy effective Lagrangian was obtained
in \cite{SW,SW2} using the symmetries of the theory as well as
certain assumptions about the nature of singularities on the modular space
of vacua. It is a challenging problem to reobtain these results
as well as to calculate
corrections to them
performing  explicit calculations in microscopic theory.
In this paper we are making another step in this direction.

To be more specific we consider N=2 SUSY SU(2) gauge theory without matter
hypermultiplets.
The theory has a flat direction along which the complex scalar field can develop
a vacuum expectation value (vev). The gauge group is broken to U(1). Thus,
low energy theory contains an abelian N=2 supermultiplet.
If we ignore high derivative/many fermions terms then the low energy
effective Lagrangian is strongly constrained by supersymmetry. The most general
form of the action \cite{SW}
\beq
S_{SW}=\int d^{4}x
\left[\int  d^{4}\theta
\frac{1}{2}\frac{\prt}{\prt A}{\cal F} (A) \bar{A}  +
\int d^{2}\theta\frac{1}{4}\frac{\prt^{2}}{\prt A^{2}}{\cal F }(A)
W_{\alpha}W^{\alpha}
+ c.c \right]
\eeq
depends on the one holomorphic  function ${\cal F} (A)$
 on the moduli
 space of vacua.
This function determines the effective coupling constant as a function of vev.
 Here we use N=1 superfield
notations, $W$ is the field strength of the N=1 vector supermultiplet, while 
$A$ is the N=1 chiral supermultiplet of U(1) theory ($a$ is the scalar
component of $A$, which develop vev).

The function ${\cal F}(A)$ was determined exactly in \cite{SW}. Its expansion
in inverse powers of  $A$  is given by
\beq
{\cal F} (A)=\frac{1}{8\pi^2}\left[A^2 log\frac{2A^2}{\Lambda^2}
- \frac{\Lambda^4}{A^2} + \cdots \right],
\eeq
where $\Lambda$ is the scale of the theory.

At large $a$ the microscopic theory is in the weak coupling and could be
studied
semiclassically.
In particular, several
 terms in the expansion of
the function ${\cal F} (A)$ in the weak coupling limit were reobtained
in a straightforward manner making explicit
calculations in the microscopic theory.

Namely,  the logarithmic term in (1.2) comes from the 
one loop contribution to the $\beta$-function. There are no higher order
perturbative corrections in N=2 supersymmetry \cite{S}
. The second term in (1.2)
comes from one instanton while ellipses stands for high order multi-instanton
corrections. The one-instanton contribution to (1.1) was first studied in
\cite{S} (see ref. \cite{AKMRV} for a review of the instanton calculus
in the SUSY theories).
 The numerical coefficient in front of it was calculated in
\cite{FP} and was shown to coincides with the prediction of the Seiberg-Witten
solution (1.2) \footnote{This calculation requires fixing of the regularization 
scheme,
both in perturbative theory and in the instanton sector. This
defines the particular scale parameter $\Lambda$. Equation (1.2)
is written down in the Pauli-Villars regularization scheme \cite{FP}.
We use the same scheme in this paper.} 
 In ref. \cite{DKM}
the two-instanton effect was studied and the coefficient in front
of $\Lambda^8/A^6$ in the expansion (1.2) was calculated in accordance
with Seiberg-Witten result.

 In this paper we continue
a detailed study of  instanton effects in the low
 energy effective theory
 making a step in another direction.
 We restrict ourselves to the one-instanton 
effects considering weak coupling limit of the theory and study
large distance expansion in the low energy Lagrangian.
 Namely, we go beyond the leading at low energies order
approximation (1.1) and present a general method to calculate
corrections to (1.1) in higher derivatives of fields.

In sect. 2 we calculate the instanton fields in terms of N=1 supermultiplets
generalizing  the method of ref. \cite{NSVZ,FS} developed for N=1 SUSY to
N=2 case. Then in sect. 3 we derive the effective instanton-induced vertex
following the general method of \cite{CDG,SVZeff,Yeff} (see also \cite{Yd=2}).
This vertex  encodes instanton-induced effects in the microscopic
theory. In sect. 4 we rewrite it in N=2 superfields and then
 consider it in the low energy limit in sect. 5.
 The leading
term appears to be the Seiberg-Witten effective action (1.1) with
function ${\cal F}(A)$ given by the second term in (1.2). We also calculate
all orders of corrections to (1.1) in the long-distance expansion.
 Sect. 6 contains our conclusions.

\section{N=2 superinstanton.}
\setcounter{equation}{0}

Consider N=2 SU(2) gauge theory. It can be described in terms of N=1
superfields by the chiral scalar multiplet
\beq
\Phi = \phi +\sqrt{2}\theta_{\alpha}\psi^{\alpha} + \theta^2 F
\eeq
and by the field strength  chiral multiplet
\beq
W^{\alpha}=-\lambda^{\alpha} + \theta^{\alpha}{\cal D } + \frac{1}{2}
(\sibsi \theta )^{\alpha}\F -\theta^2 i\nd{D}^{\alpha\dot{\alpha}}\bar{\lambda}
_{\dot{\alpha}}
\eeq
in Wess-Zumino gauge.
Both fields in (2.1) and (2.2) are in the adjoint representation of SU(2)
and carry the colour index $a=1, 2,3$. Sigma matrices in (2.2) are
defined as follows $\sigma_{\mu}^{\al \dal}=(1,-i\tau^{a}),\; 
\bar{\sigma}_{\mu \dal \al}=(1,i\tau^{a})$ , $\tau^a$ being the Pauli
 matrices. 
The action of the model reads
\begin{eqnarray}
S &  = &  \frac{1}{g^2}\int d^4 x \left[
\int d^2 \theta d^2 \bar{\theta}\bar{\Phi}^a (e^{-2V_{g}}\Phi)^a +
\frac{1}{4g^2}\int d^2 \theta W_{\alpha}^a W^{\alpha a}
\right. \nonumber \\
 & + & \left. \frac{1}{4g^2}\int d^2 \bar{\theta}
\bar{ W}^{\dot{\alpha}a}\bar{W}_{\dot
{\alpha}}^{ a} \right] .
\end{eqnarray}
N=2 SUSY assumes the existence of global $ SU(2)_{R}$ group. Fermions
$\lambda,\;\psi$ form a doublets under the action of this group, while
gauge and scalar fields are singlets.
 
As we already mentioned above, this theory has a flat direction in the 
scalar potential. Namely, the condition of vanishing of the D-term
is $[\bar{\phi},\phi]=0$ (troughout the paper we use both the matrix as well as
the vector notations for fields in the adjoint representation of the colour
group, say $\phi=\phi^a\tau^a /2$). As
a solution for the vacuum moduli space we use $\phi=v\tau^3 /2$ following
\cite{SW}. Thus, the colour group is broken down to U(1) and  light
fields which appear in the low energy theory (1.1) are
\beq
A=\Phi^3,\; W=W^3
\eeq
For large $v\gg \Lambda$ the effective coupling is small and the theory
can be studied in the semiclassical approximation.
As we mentioned in the Introduction the perturbative 
contribution to the effective coupling
\beq
\frac{1}{g^{2}_{eff}}=\frac{\prt^2 {\cal F}}{\prt A^2}
\eeq
is exactly given by one loop result (first term in (1.2)) so we are left
only with
instanton effects.

 In the reminder of this section we present different components of the 
instanton fields in the theory (2.1) and arrange them into N=1
supermultiplets.

Let us start with the boson fields first. The gauge potential is the 
original BPST instanton \cite{BPST} written down in the singular
gauge
\beq
A^{I}_{\mu}=\eta^a_{\mu\nu}\rho^2\frac{u\tau^a \bar{u}y_{\nu}}{y^4 H},
\eeq
where $y=x-x_{0}$ and
\beq
H=1+\frac{\rho^2}{y^2}.
\eeq
Here $x_{0\mu},\rho$ and $u$ are the instanton center, its size and orientation
matrix $u^{\alpha\dot{\alpha}}=\sigma^{\alpha\dot{\alpha}}_{\mu}u_{\mu}$,
 $u_{\mu}^2=1$.
 In order to construct
the chiral field $W$ in (2.2) we need the expression for the gauge 
field strength
\beq
\frac{1}{2}(\sibsi )^{\alpha}_{\beta}\F^{Ia}
=-8i\rho^2\frac{(y\bar{u}\tau^au\bar{y})^{
\alpha}_{\beta}}{y^6 H^2},
\eeq
where matrix $y^{\alpha\dot{\alpha}}=\sigma^{\alpha\dot{\alpha}}_{\mu}y_{\mu}$.

The scalar component of the instanton field in the Higgs vacuum is 
given by \cite{H}
\beq
\phi_{I}=\frac{\tau^3}{2}\frac{v}{H}
\eeq
Strictly speaking the configuration (2.6), (2.9) is not an exact solution of 
equations of motion in the Higgs vacuum.
The scalar field (2.9) satisfies the equation $D^2\phi_{I}=0$, but the gauge
field in (2.6) do not satisfy the equation of motion with the source term
induced by scalars. This term becomes important at large distances
$y^2\geq M_{W}^{-2}$, where $M^{2}_{W}\sim v^2$ is the mass of the W boson.
 The solution (2.6), (2.9) can be viewed as a constraind
instanton \cite{A}. It is given by (2.6), (2.9) at short distances, whereas
has an exponential fall off at large distances for heavy components.
The action on fields (2.6), (2.9) is
\beq
S_{bos}^{I}= \frac{8\pi^2}{g^2}+\frac{ 4\pi^2}{g^2} |v|^2\rho^2.
\eeq
This means that the integral over the instanton size  is convergent
at large $\rho$ due to the second term in (2.10) and dominated at
\beq
\rho^2 \sim \frac{g^2}{|v|^2}.
\eeq
We can look at bosonic instanton fields as given by (2.6), (2.9) at all
distances up to $y^2\leq \mu^{-2}$ where scale $\mu$ is within the bounds
$\rho^2\ll \mu^{-2}\ll M_{W}^{-2}$ 
. Because of (2.11) we have some range of validity of eqs. (2.6), (2.9)
even in the large distance limit $y^2\gg \rho^2$ we are interested in in this
 paper. To go beyond that at very large distances $y^2\gg v^{-2}$
(which is essential
to derive the low energy Lagrangian (1.1)) we have to take into account
the mass term  for heavy fields in (2.3), while light components 
are still given
by the leading terms of large distance expansion in 
 eqs. (2.6), (2.9). We make this extrapolation in sect. 5.

Let us now discuss fermion zero modes of instanton. There are eight
of them, four for $\lambda$ and four for $\psi$, if $v=0$. Gaugino ones
are given by \cite{ADS,NSVZ}
\beq
\lambda^{a \alpha}
=8i\rho^2\frac{(y\bar{u}\tau^au\bar{y}(\alpha -y\bar{\beta}))^{
\alpha}}{y^6 H^2},
\eeq
Here we introduce two Grassmann parameters $\alpha^{\alpha}$ which
parametrize supersymmetric zero modes as well as  $\bar{\beta}_{\dot{\alpha}}$
which parametrize superconformal ones.
To get the $\psi$ modes we use the $SU(2)_{R}$ group which interchanges
$\lambda$ and $\psi$. We have
\beq
\psi^{a \alpha}
=-8i\rho^2\frac{(y\bar{u}\tau^au\bar{y}(\zeta -y\bar{\omega}))^{
\alpha}}{y^6 H^2},
\eeq
where we introduce another set of Grassmann parameters $\zeta^{\alpha}$ and 
$\bar{\omega_{\dot{\alpha}}}$.

Once $v\neq 0$ superconformal modes in (2.12), (2.13) are lifted due to the 
conformal symmetry breaking. Nevertheless we still keep the integration over 
corresponding parameters $\bar{\beta}$ and $\bar{\omega}$ in the instanton 
measure following \cite{ADS, NSVZ}. These are fermionic counterpart of the 
integration over the instanton size $\rho$.

Now let us rewrite components of instanton field in terms of N=1 supermultiplets 
using the general method of ref \cite{NSVZ,FS}. Note first, that under the N=1 
SUSY transformation components of vector multiplet (2.6) and (2.12) transform 
into the same expressions up to a gauge transformations with collective 
coordinates redefined as follows:
\begin{eqnarray}
x'_{0\mu} & = & x_{0\mu} - 2i(\bar{\epsilon}\bar{\sigma}_{\mu}\alpha), 
 \nonumber  \\
\alpha' & = & \alpha - \epsilon, \nonumber  \\  
\rho'^{2} & = & \rho ^{2}(1 + 4i\bar{\beta}\bar{\epsilon}), \nonumber  \\
\bar{\beta'} & = & \bar{\beta}(1 + 4i\bar{\beta}\bar{\epsilon}), \nonumber  \\
u'^{\alpha\dot{\alpha}} & = & u^{\alpha\dot{\beta}}[1 - \delta^{\dot{\alpha}}
_{\dot{\beta}}2i\bar{\beta}\bar{\epsilon} - 
4i\bar{\beta}_{\dot{\beta}}\bar{\epsilon}^{\dot{\alpha}}],
\end{eqnarray}
where $\epsilon$ and $\bar{\epsilon}$ are parameters of the N=1 SUSY
transformation.

These expressions were obtained in \cite{NSVZ,Y88,Y90} for 
 N=1 SUSY gauge theory and stay untouched for the vector supermultiplet in N=2 
SUSY theory. To derive transformation laws for $ \zeta $ and $\bar{\omega}$ let 
us rewrite the supersymmetric mode in (2.13) as
\beq
   \psi^{\al}_{SS} = \frac{1}{2}(\sibsi\zeta)\F^I 
\eeq
and superconformal one as 
\beq
   \psi^{\al}_{SC} = - i\sqrt{2}\nd{D}^{\al\dal}\phi^{I}\bar{\eta}_{\dal},
\eeq
where $\F^I $ and $\phi^I $ are given by (2.8), (2.9). Here we introduce a new 
collective coordinate $\bar{\eta}$ instead of $\bar{\omega}$ :
\beq
   \bar{\omega}_{\dal} = \frac{v}{2\sqrt{2}}(\bar{u}\tau^{3}u\bar{\eta})_{\dal}
\eeq

Using (2.15), (2.16) it is easy to see that N=1 SUSY transformations lead to the 
following transformation law for $\zeta$, $\bar{\eta}$:
\begin{eqnarray}
   \zeta' & = & \zeta  \nonumber  \\
   \bar{\eta}' & = & \bar{\eta} - \bep + 4i\bbt(\bar{\eta}\bep)
\end{eqnarray} 

Now we are ready to construct instanton supermultiplets. Fields $\Phi$ and $W$
 are chiral superfields, therefore, they should be invariant (up to a gauge 
transformation) under the SUSY transformation of coordinates
\begin{eqnarray}
  x'_{L\mu} & = & x_{L\mu} - 2i(\bep\bsi_{\mu}\theta)  \nonumber  \\
 \theta' & = & \theta - \ep
\end{eqnarray}
followed by transformation of the collective coordinates (2.14), (2.18). Here 
$x_{L\mu}$ and $\theta$ are arguments of chiral field
\beq
    x_{L\mu}  =  x_{\mu}+ i(\bar{\theta}\bsi_{\mu}\theta) 
\eeq

Comparing (2.14) with (2.19) we see that instanton center transforms like a 
left-handed coordinate, while $\al$ transforms like $\tht$. To complete this 
analogy showing that instanton collective coordinates can be interpreted as
the usial superspace coordinates,
 let us introduce
\beq
   \bar{\eta}_{1}= \frac{\bar{\eta}}{1+4i\bar{\bt}\bar{\eta}},
\eeq
which gets shifted
\beq
   \bar{\eta}_{1}'= \bar{\eta}_{1}-\bep
\eeq
under the SUSY transformations \cite{Y90}. Eq. (2.22) shows that
$\bar{\eta}_{1}$ transforms like $\bar{\theta}$.

The invariance of fields $\Phi$ and W means, that they could depend only on 
certain SUSY invariant combinations of coordinates $x_{L},\;\tht $ and 
collective coordinates (cf. \cite{NSVZ}). Using (2.19) as well as (2.14) and 
(2.18) we can
 construct this invariants. The invariant distance is
\beq
   y^{inv}_{L\mu} = x_{L\mu} - x_{0\mu}+ 2i(\bar{\eta}_{1}\bsi_{\mu}[\tht+
 \al]),
\eeq
where $\tht+ \al$ is invariant by itself. The invariant size of instanton is 
\cite{NSVZ}
\beq
  \rho^{2}_{inv}=\rho^{2}(1+4i\bbt\bet), 
\eeq
while invariant $\bbt$ is \cite{Y90}
\beq
  \bbt_{inv}=\bbt(1+4i\bbt\bet). 
\eeq

The choice of the invariant distance in (2.23) is slightly different from those 
in \cite{NSVZ}. The one in (2.23) looks like invariant distance between ordinary 
left-handed coordinates. We used this analogy in the next section. To 
construct fields $\Phi^{a}$, $W^{\al a}$ (rather then gauge invariant 
combinations $W^{a}W^{a}$ and $\Phi^{a}\Phi^{a}$) we also need a notion of the 
invariant orientation. It is given by
\beq
  u^{\al\dal}_{inv}= u^{\al\dbt}exp[-4i\bbt_{\dbt}\bet^{\dal} -
 2i\dl^{\dal}_{\dbt}(\bet\bbt)]
\eeq

Now we can write down the answer for vector field (cf. \cite{NSVZ})
\beq
  W^{\al}= \frac{1}{2}(\sibsi[\tht + \al - y^{inv}_{L}\bbt_{inv}])^{\al}
      \F^{I}(y^{inv}_{L},\rho_{inv},u_{inv}),
\eeq
where $\F^{I}$ is given by (2.8). It is a straightforward calculation to check 
that components of (2.27) are given by (2.8), (2.12) up to a gauge 
transformation. Note, that extra terms which comes from expansion of 
$\F^{I}(y^{inv}_{L})$ in the deviation of $y^{inv}_{L}$ from $y_{L}$ (see 
(2.23)) are zero by the use of equation of motion. A useful formula to make this 
check is 
$$
 \frac{y^{inv}_{L}}{\rho_{inv}}= \frac{y_{L}-4i[\tht+\al-y_{L}\bbt]\bet}{\rho}
\bar{u}u_{inv}.
$$

The similar procedure gives for the scalar chiral field
$$
 \Phi = \frac{1}{\sqrt{2}}([\tht + \al - y^{inv}_{L}\bbt_{inv}]
     \sibsi \zeta ) \F^{I}(y^{inv}_{L},\rho_{inv},u_{inv})
$$
\beq
 + U_{1}\phi_{I}
    (y^{inv}_{L},\rho_{inv})U^{-1}_{1},
\eeq
where $U_{1}$ is the unitary rotation.
\beq
 U_{1} = exp[2A^{I}_{\mu}(\bet\bsi_{\mu}[\tht + \al - y_{L}\bbt])]
\eeq
This rotation could be ignored in large distance limit $x^{2}>>\rho^{2}$. Again 
one can check that the components of (2.28) are given by (2.9), (2.13).

It is important to note that supersymmetric zero mode of $\psi$ (2.15) comes
 from the first term in (2.28) while the superconformal one comes from the
 expansion of the second one. This is the way N=2 SUSY works. Roughly 
speaking, the 
first term in (2.28) is an N=2 supersymmetrization of the original BPST 
instanton in (2.8) (action of N=1 SUSY gives $\la_{SS}$ in (2.12), 
then the action of $SU(2)_{R}$ gives $\psi_{SS}$ in (2.13). The second term
 in (2.28) is the supersymmetrization of the t'Hooft solution (2.9) which
 seems unrelated to (2.8). They ``meet together'' in (2.28) as the first term 
produces supersymmetric zero mode (2.15), while the second one produces the 
superconformal mode (2.16). To put it another way, the coefficient 
$\bar{\omega}$ in front of $\psi_{SC}$ in (2.13) is related to N=2 
superconformal transformation of BPST instanton (2.8), while parameter $\bet$ is 
related to N=1 SUSY transformation of the t'Hooft solution (2.9). Eq. (2.17) is 
the equivalence between both.

Now let us turn to anti-chiral fields
\beq
  \bPh= \bph + \sqrt{2}\bth^{\dal}\bar{\psi}_{\dal}+\bth^{2}F^{2}
\eeq
and
\beq
 \bar{W}_{\dal}= -\bla_{\dal} + \bth_{\dal}{\cal D}- \frac{1}{2}(\bsisi\bth)
   _{\dal}\F - \bth^{2}(i\nd{\bar{D}}\la)_{\dal}
\eeq
which depend on $\bth$ and right-handed coordinate
\beq
 x_{R\mu}= x_{\mu}-i(\bth\bar{\si}_{\mu}\tht)
\eeq
If the vev of the scalar field were zero there would be no anti-chiral fields. 
Instanton would be self-dual, chiral object. Once $v\neq0$ this is no longer 
true \cite{ADS}. The $\bph$ component is given by complex conjugate of the 
t'Hooft solution (2.9)
\beq
  \bph_{I} = \frac{\tau_{3}}{2}\frac{\bar{v}}{H}
\eeq
Then N=1 SUSY transformation generates
\beq
 \dl\bar{\psi}_{\dal} = i\sqrt{2}\bsi_{\mu\dal\al}\ep^{\al}D_{\mu}\bph_{I} 
\eeq
This field becomes a nontrivial solution of equations of motion
\beq
 i\nd{D}\bar{\psi} - i\sqrt{2}[\la_{SS},\bph_{I}] = 0
\eeq
once $\bph \neq 0$. From (2.34),(2.35)we have
\beq
 \bar{\psi}_{\dal} =  i\sqrt{2}\bsi_{\mu\dal\al}\al^{\al}D_{\mu}\bph_{I}
\eeq

Note, that according to (2.35) we parametrize $\bar{\psi}$ mode in (2.36) by 
parameter $\al$, which appears in the expression for supersymmetric gaugino zero 
mode (2.12). The $SU(2)_{R}$ symmetry then gives for $\bar{\la}$
\beq
 \bar{\la}_{\dal}= -i\sqrt{2}\nd{\bar{D}}_{\dal\al}\phi_{I}\zeta^{\al}
\eeq

Now let us construct anti-chiral supermultiplets. To do this we should introduce 
the right-handed invariant distance. Recalling that
\begin{eqnarray}
  x'_{R\mu} & = & x_{R\mu} + 2i(\bth\bsi_{\mu}\ep)   \nonumber  \\
   \bth' & = & \bth -\bep
\end{eqnarray}
and taking into account (2.14) one finds
\beq      
  y^{inv}_{R\mu} = x_{R \mu} - x_{0 \mu}  -2i(\bth\bsi_{\mu}\al)
\eeq
Using this invariant distance as well as $\rho_{inv}$ and $u_{inv}$ (2.24), 
(2.26) we make an obvious guess
\beq      
 \bPh = U\bph_{I}(y^{inv}_{R}, \rho_{inv})U^{-1}
\eeq
for scalar multiplet and
\beq      
  \bar{W}_{\dal} = i\sqrt{2}U\nd{\bar{D}}_{\dal\al}\phi_{I}(y^{inv}_{R}, 
\rho_{inv}, u_{inv})U^{-1}\zeta^{\al}
\eeq
for the vector one. Here
\beq      
 U = exp[-2A^{I}_{\mu}(\bth\bsi_{\mu}\al)]
\eeq
is the unitary rotation which transforms
 ordinary derivatives into covariant ones in the expansion of these expressions 
in the deviation $(y^{inv}_{R} - y_{R})$
(see (2.39)). This rotation is inessential in the large distance limit.

Eqs.(2.27) and (2.28), as well as (2.40) and (2.41), are our manifestly N=1 
supersymmetric results for instanton supermultiplets.

\section{Instanton-induced effective vertex}
\setcounter{equation}{0}

In this section we are going to derive the effective vertex which once added to 
the tree level Lagrangian (2.1) reproduces instanton contributions to all 
correlation functions in the framework of a perturbation theory.

Let us first recall the notion of this vertex in non-supersymmetric theories. 
Consider pure gauge theory.

The instanton induced effective vertex suggested by Callan, Dashen and 
Gross a long 
time ago \cite{CDG} (see also \cite{SVZeff}) has the form
\beq      
 V^{CDG}_{I} = -c\int d^{4}x\frac{d\rho}{\rho^{5}}\frac{d^{3}u}{2\pi^{2}}
    (\rho\La)^{b}exp[\frac{\pi^{2}}{g^{2}}i\rho^{2}Tr(\bsisi\bar{u}\tau^{a}u
    )\F^{a}]
\eeq
where $b$ is the first coefficient of the $\bt$-function, $c$ is a number.
 To derive it let us consider the n-point correlation function
\beq      
  <A_{\mu 1}(x_{1}),\ldots,A_{\mu n}(x_{n})>
\eeq
in the instanton background in the large distance limit $(x_{i} - x_{0})^{2}
  \gg \rho^{2}, i=1,\ldots,n$. Here the double index $\mu i$ denotes the 
set of indices $\mu_{1}\ldots \mu_{n}$ 
On one hand, in the leading semiclassical approximation $(g^{2}\ll1)$ the 
correlation function (3.2) is given by the product of the classical expressions 
(2.6) for each $A_{\mu i}(x_{i})$
\beq      
  \prod_{i=1}^{n} \eta^{ai}_{\mu i\nu i} \rho^{2}\frac{u\tau^{ai}\bar{u}}
  {(x_{i}-x_{0})^{4}}(x_{i}-x_{0})_{\nu 
i}[1+O(\frac{\rho^{2}}{(x_{i}-x_{0})^{2}})]
\eeq
Note, that we expand (2.6) in the limit $(x_{i}-x_{0})\gg\rho^{2}$.

 On the other hand, to get the same expression using (3.1), let us add (3.1) to 
the classical action 
 and than expand $e^{-V}=\sum_{k}\frac{1}{k!}(-V^{k})$.
Each term in this expansion corresponds to the k-instanton contribution. Now we 
concentrate on k=1. We have for (3.2)
\beq      
  <A_{\mu 1}(x_{1}),\ldots,A_{\mu n}(x_{n})V^{CDG}_{I}>_{pert},
\eeq
where average means the integration over quantum fluctuations around
 perturbative
vacuum. Expanding the exponent in (3.1) n times and using
\beq      
  <A_{\mu}^{a}(x),F^{b}_{\al\bt}(x_{0})>_{pert}=\frac{2g^{2}}{(2\pi)^{2}}
  \dl^{ab}\frac{(\dl_{\mu\bt}y_{\al}-\dl_{\mu\al}y_{\bt})}{y^{4}}
\eeq
in the large distance limit we arrive at (3.3).

Actually in this paper we will use the effective vertex method outlined above 
only in the large distance limit $(x_{i}-x_{0})^{2}\gg\rho^{2}$. However, it is 
worth noting that it is much more powerful. Namely, the effective vertex in 
(3.1) reproduces not only large distance behaviour of the classical instanton 
field in (2.6) but $\rho^{2}/y^{2}$ corrections as well \cite{Z}. These 
corrections comes from loop graphs. We can write symbolically the 1-point 
correlation function in (3.4)
\beq      
  <A(x),V^{CDG}_{I}>_{pert}=\frac{1}{g^{2}}<A,F>_{pert}+\frac{1}{2!g^{4}}
<A,F,F> _{pert}+\cdot .
\eeq
The first term here gives the leading order contribution (3.5) we considered 
above. The others are nonzero provided interaction vertices from the classical 
Lagrangian are inserted. For example, the second term in the r.h.s. of (3.6) 
requires the insertion of the three-gluon vertex. However, these corrections are 
not suppressed by $g^{2}$ \cite{Z}. The extra powers of $g^{2}$ are canceled 
against powers $1/g^{2}$ appeared in the expansion (3.6). In fact, these higher 
order loop graphs give proper expansion of instanton field (2.6) in powers of 
$\rho^{2}/y^{2}$ \cite{Z}.

It is worth note also that instanton-induced vertices of type (3.1) describe not 
only single instanton effects but multi-instanton interactions as well. In 
particular, instanton-anti-instanton interaction comes from
\beq      
  <V_{I},V_{\bar{I}}>_{pert},
\eeq
where $V_{\bar{I}}$ for anti-instanton is given by the similar to (3.1) formula 
with the replacement $\si \leftrightarrow \bsi$,  $u \leftrightarrow \bar{u}$. 
The above statement was checked in the large distance limit for pure gauge 
theory \cite{CDG} as well as for gauge-Higgs theory \cite{Yeff} and N=1 SUSY QCD 
\cite{Y90}.

Let us now make our theory more complicated adding Higgs fields in the adjoint 
representation.

The effective vertex takes the form
\beq      
  V^{GH}_{I} = c_{GH}\int d^{4}x\frac{d\rho}{\rho_{5}}(\rho\La)^{b}
  \frac{d^{3}u}{2\pi^{2}}exp[-\frac{4\pi^{2}}{g^{2}}\rho^{2}\bph^{a}\phi^{a}
  + \frac{\pi^{2}}{g^{2}}i\rho^{2}Tr(\bsisi\bar{u}\tau^{a}u)\F^{a}],
\eeq
where $c_{GH}$ is another calculable constant.

The similar vertex for Higgs fields in the fundamental representation was 
obtained in \cite{Yeff}. To derive (3.8) let us expand $\phi$ around its vev.
 We have
\beq      
  \bph^{a}\phi^{a} = |v|^{2} +\bar{v}\dl\phi^{3} +\dl\bph^{3}v +
   \dl\phi^{a}\dl\phi^{a}.
\eeq
It is sufficient to consider only correlation functions of Higgs fields
\beq      
  <\phi^{a1}(x_{1}),\ldots,\phi^{an}(x_{n}),\bph^{b1}(z_{1}),\ldots,\bph^{bk}
    (z_{k})>,
\eeq
because 
 the factor associated with gauge fields in (3.8) stays untouched as compared
 with (3.1) and produces the same correlation functions (3.3) of gauge fields.

First note, that $|v|^{2}$ term in (3.9) when inserted in (3.8) gives correctly 
the piece of the instanton action (2.10) associated with scalars. Consider now 
the second and the third terms in r.h.s. of (3.9) linear in quantum 
fluctuations. Expanding $exp-V^{GH}_{I}$ in $\dl\bph v$ and $\bar{v}\dl\phi$ and 
using propagation function for scalar field
\beq      
  <\phi^{a}(x),\dl\bph^{3}(x_{0})>=\dl^{a3}\frac{g^{2}}{(2\pi)^{2}}\frac{1}
   {y^{2}}
\eeq
one gets for the correlation function (3.10)
\beq      
 \prod_{i=1}^{n}\dl^{ai3}v[1-\frac{\rho^{2}}{(x_{i}-x_{0})^{2}}]\prod_{j=1}^{k}
    \dl^{bj3}\bar{v}[1-\frac{\rho^{2}}{(z_{j}-x_{0})^{2}}]
\eeq
This is exactly the same expression one gets in large distance limit 
substituting classical t'Hooft solutions (2.9), (2.33) into (3.10).

As for the last term in r.h.s. of (3.9), quadratic in fluctuations, it 
corresponds to quantum corrections and could be restored only if quantum 
propagators in
the external instanton field are taken into account. An explicit check on this 
term was made in \cite{LVV} for Higgs fields in the fundamental representation. 
Note, that the promotion
\beq      
 |v|^{2} + \bar{v}\dl\phi^{3}+ \dl\bph^{3}v \rightarrow \bph\phi
\eeq
which comes in a non-trivial way from quantum corrections in the language of 
correlation functions, appears as an obvious guess in the effective vertex 
approach. The reason is that effective Lagrangian cannot depend on vev of a 
given field (that would be signal for an explicit symmetry breaking rather than
 spontaneous). Effective Lagrangian ``does not know'' whether the spontaneous 
symmetry breaking occurs or not, and could depends only on fields. Therefore, 
the promotion 
(3.13) we made in (3.8) looks as a natural generalization compatible with
 symmetries of the theory. We use this rather powerful idea throughout the 
paper.

Now let us consider our N=2 SUSY gauge model. In fact, the instanton-induced 
vertex in this model is a straightforward N=2 supersymmetrization of (3.8). 
However, we will derive it first in terms of N=1 supermultiplets using results 
of the previous section. First we write down the result and then check it using 
the same method as for pure gauge and gauge-Higgs theories. Our result is
\begin{eqnarray}
  V^{N=2}_{I} & = & -c_{N=2}\La^{4}\int d^{4}x_{0}\frac{d\rho}{\rho}
   \frac{d^{3}u_{inv}}{2\pi^{2}}d^{2}\al d^{2}\zeta d^{2}\bbt_{inv}
   d^{2}\bet_{1}\frac{1}{\Phi^{a2}}   \nonumber  \\
 & &  exp \left[ - \frac{4\pi^{2}}{g^{2}}\rho^{2}_{inv}\bPh^{a}\Phi^{a}
    -\frac{4\sqrt{2}\pi^{2}}{g^{2}}\rho^{2}_{inv}\bPh^{a}W^{a}_{\al}
    \zeta^{\al} \right. \nonumber  \\
          & + & \frac{2\pi^{2}}{g^{2}}i\rho^{2}_{inv}\bar{\nabla}_{\dbt}
    (\bar{W}^{a}\bar{u}_{inv}\tau^{a}u_{inv})^{\dbt} - \frac{16\pi^{2}}{g^{2}}
\rho^{2}_{inv}(\bar{W}^{a}\bar{u}_{inv}\tau^{a}u_{inv}\bbt_{inv}) \nonumber \\ 
 & + &  \frac{2\sqrt{2}\pi^{2}}{g^{2}}
   \rho^{2}_{inv}(\zeta\nd{D}\bar{u}_{inv}\tau^{a}u_{inv}\bar{\nabla}\bPh^{a}) 
\nonumber \\
& - & \left.\frac{16\sqrt{2}\pi^{2}}{g^{2}}i
   \rho^{2}_{inv}(\zeta\nd{D}\bar{u}_{inv}\tau^{a}u_{inv}\bbt_{inv})\bPh^{a}  
\right],
\end{eqnarray}
where supercovariant derivatives defined in the usual way
\begin{eqnarray}
\bar{\nabla}_{\dal} & = & \frac{\prt}{\prt\bth^{\dal}} \nonumber  \\
\nabla^{\al} & = & \frac{\prt}{\prt\tht_{\al}} - 2i\nd{\prt}^{\al\dal}
  \bth_{\dal}
\end{eqnarray}
when acting on functions of left-handed coordinate $x_{0}$.

Here N=1 superfields are understood as the following chiral functions
\begin{eqnarray}
\Phi & = & \Phi (x_{0}, \tht=-\al) \nonumber  \\
  W & = & W(x_{0}, \tht=-\al), 
\end{eqnarray}
as well as anti-chiral functions
\begin{eqnarray}
\bPh & = & \bPh (x^{R}_{0}, \bth=-\bet_{1}) \nonumber  \\
 \bar{W} & = & \bar{W}(x^{R}_{0}, \bth=-\bet_{1}),
\end{eqnarray}
where right-handed instanton center is 
\beq      
 x^{R}_{0\mu} = x_{0\mu}-2i(\bet_{1}\bsi_{\mu}\al).
\eeq
The numerical constant $c_{N=2}$ will be fixed below.

Now let us check the result (3.14). First note that deriving the vertex (3.1) 
for gauge theory in the semiclassical approximation we could restrict ourselves
 to checking only 1-point correlation function instead of (3.2). The reason is 
that exponential nature of (3.1) will reproduce then the result (3.3) for 
n-point correlation function with correct combinatorics automatically. In a 
similar way, 
for the gauge-Higgs theory we could 
make a check for the correlation function (3.10) with n=1, k=1 only due to the 
same reason.

Therefore, to  check the exponential factor in (3.14) it is sufficient to 
consider the following correlation function
\beq      
 <\Phi^{a}(x^{L}_{1},\tht_{1}),\bPh^{b}(x^{R}_{2},\bth_{2}),
   \bar{W}^{d}_{\dal}(x^{R}_{3},\bth_{3}), W^{\al c}(x^{L}_{4},\tht_{4})>
\eeq
in the instanton background. Let us plug $exp(-V^{N=2}_{I})$ into (3.19) and
 keep only the one-instanton contribution $(-V^{N=2}_{I})$. Now consider the 
first term in the square brackets in (3.14). Expanding $\bPh$ and $\Phi$ around 
their expactation values, we have
\beq      
 \bPh^{a}\Phi^{a} = |v|^{2} + \bar{v}\dl\Phi^{3}+\dl\bPh^{3}v+
      O(\dl\bPh\dl\Phi),
\eeq
where we dropped out the term quadratic in quantum fluctuantions.

Expanding the exponent in (3.14) in powers of $\dl\bPh^{3}v$ and contracting the 
 field $\Phi^{a}(x^{L}_{1},\tht_{1})$ with $\dl\bPh^{3}(x^{R}_{0},
\bth=-\bet_{1})$, using the propagation function for chiral fields
\beq      
 <\Phi^{a}(x^{L}_{1},\tht_{1}),\delta \bPh^{b}(x^{R}_{0},\bth)> = 
  \frac{\dl^{ab}g^{2}}{(2\pi)^{2}}e^{-2i\bth\nd{\bar{\prt}}\tht_{1}}
   \frac{1}{(x^{L}_{1}- x^{R}_{0})^{2}}
\eeq
one gets for the first factor in (3.19)
\beq      
 \dl^{a3}{v}[1 - \frac{\rho^{2}_{inv}}{(x^{L}_{1\mu}-x_{0\mu}
  +2i\bet_{1}\bsi_{\mu}[\tht_{1}+\al])^{2}}]
\eeq
which is nothing else then the second term in the expression (2.28) for scalar 
supermultiplet $\Phi(x^{L}_{1},\tht_{1})$ in the large distance limit. The first 
term we will recover later on. The invariant distance in the (3.22) comes as 
follows. The shift in $2i\bet_{1}\bsi_{\mu}\tht_{1}$ comes from the exponential 
in (3.21) at $\bth=-\bet_{1}$ (see (3.17)), while the shift 
$2i\bet_{1}\bsi_{\mu}\al$ appears when we substitute in (3.21) the right-handed 
instanton center (3.18).

Now consider $\bar{v}\dl\Phi^{3}$ term in the exponent in (3.14). Contracting it 
with the field $\bPh(x^{R}_{2},\bth_{2})$ (3.19) we get 
\beq      
 \dl^{b3}\bar{v}[1 - \frac{\rho^{2}_{inv}}{(x^{R}_{2\mu}-x_{0\mu}
  -2i\bth_{2}\bsi_{\mu}\al)^{2}}].
\eeq
This is  the  expression (2.40) for $\Phi(x^{R}_{2},\bth_{2})$
in the large distance limit. Now to get the first term in the exponent in (3.14) 
we promote (3.20) to the full expression $\bPh^{a}\Phi^{a}$. Although this term 
respects N=1 SUSY it is not complitely gauge invariant due to the shift
 (3.18). Therefore, we would expect that the correct promotion of (3.20) in 
(3.14) is $\bPh^{a}(e^{-2Vg}\Phi)^{a}$, instead of $\bPh^{a}\Phi^{a}$. This 
effect, however, is beyond our control in the large distance limit. 

The next step is to look at the second term in the exponent in (3.14). As usual 
we are going to recover it only in semiclassical approximation
\beq      
 \bPh^{a}W^{a}=\bar{v}W^{3} + O(\dl\bPh^{a}W^{a})
\eeq

Contracting $\bar{v}W^{3}$ term in the exponent in (3.14) with field 
$\bar{W}(x^{R}_{3},\bth_{3})$ and making use of the propagation function of the 
W-field at large distances
\beq      
 <W^{a\al}(x^{L},\tht),\bar{W}^{b\dal}(x'^{R},\bth)> =
  -\frac{ig^{2}}{2\pi^{2}}\dl^{ab}e^{-2i\bth\nd{\bar{\prt}}\theta}\frac{(x^{L}
  -x'^{R})^{\al\dal}}{(x^{L}-x'^{R})^{4}}
\eeq
we get 
\beq      
 4\sqrt{2}i\rho^{2}_{inv}\bar{v}\frac{[(\bar{x}^{R}_{3}-\bar{x}_{0})_{\dal\al}
  -4i\bth_{3\dal}\al_{\al}]}{(x^{R}_{3}-x_{0}-2i\bth_{3}\bsi_{\mu}\al)^{4}}
\eeq

This is exactly what we need for the  field $\bar{W}_{\dal}(x^{R}_{3},\bth_{3})$ 
(2.41) in the large distance limit.

Continuing this process one can check in the same way that the third and the 
fourth terms in the exponent in (3.14) give us large distance limit of the 
expression (2.27) for the last term $W(x^{L}_{4},\tht_{4})$ in (3.19). Finally 
two last terms in the exponent in (3.14) give the missing first term in (2.28) 
for the field $\Phi(x_{1},\tht_{1})$ in (3.19). Again note, that to get 
completely gauge invariant effective vertex we should insert $e^{-2Vg}$ in all 
terms containing anti-chiral field as a function of a shifted argument (3.18).

Now let us discuss the preexponential factor in (3.14). The instanton measure in 
SUSY theories is well-known (see, for example \cite{C, FP}). The factor
 associated with the integral over bosonic collective coordinates reads \cite 
{H}
\beq      
2^{10}\pi^{6}M^{8}\rho^{8}e^{-\frac{8\qpi}{\qg}}\fd x\frac{d\rho}{\rho^{5}}
     \frac{\td u}{2\qpi}.
\eeq
Here M is the UV cutoff. The eighth power of $M\rho$ corresponds to the eight 
boson zero modes. To get the full instanton measure we have to take into account 
the normalization of fermion zero modes. Using eqs. (2.12), (2.15) and (2.16) we 
get
\beq      
  \frac{1}{16\qpi M}\qd\al\frac{1}{16\qpi M}\qd\zeta \frac{1}
{32\qpi\rho^{2} M}\qd\bbt\frac{1}{4\qpi\qrho v^{2} M}\qd\bet
\eeq
Here the numerical factor in front of each Grassmann integration accounts for 
the normalization of the corresponding fermion zero mode. Combining (3.27) and 
(3.28) together we have
\beq      
\frac{1}{32\qpi}\frac{\La^{4}}{v^{2}}\fd x\frac{d\rho}{\rho}\frac{d^{3} 
u_{inv}}{2\qpi}\qd\al\qd\zeta\qd\bbt_{inv}\qd\bet_{1}
\eeq
where $\La^{4}=M^{4}exp-8\qpi/g^{2}$ in the Pauli-Villars regularization scheme. 
We also change variables from u to $u_{inv}$ using (2.26) and from $\bbt$,
 $\bet$ to $\bbt_{inv}$, $\bet_{1}$ using $\qd\bbt\qd\bet = 
\qd\bbt_{inv}\qd\bet_{1}$ (see(2.21),(2.25)). The instanton measure in (3.29) is 
manifestly N=1 SUSY invariant. \footnote{ This is clear because we can change 
variables
 $d\rho/\rho\rightarrow d\rho_{inv}/\rho_{inv}$ in (3.29) (or (3.14)). We have 
not done this explicitly here because the integral over $\rho$ is 
logarithmically divergent at small $\rho$ and will be performed below in a more 
careful way (see also \cite {NSVZ}).}

Now we need only one final step to get the prefactor in (3.14). As we explained 
above the effective Lagrangian could not depend on the vev $v$ of the field 
$\Phi$. Therefore, we make the obvious promotion 
\beq      
\frac{1}{v^{2}}\rightarrow \frac{1}{\Phi^{a2}}
\eeq 
in the instanton measure (3.29) similar to what we made in the exponent. This 
accounts for quantum corrections which are beyond our control in the 
semiclassical approximation.

Thus we recover the effective vertex (3.14).
 The numerical coefficient $c_{N=2}$
can be read off from (3.29):
\beq      
c_{N=2}=\frac{1}{32\qpi}.
\eeq 

Let us note in the conclusion of this section that the promotion
\beq      
    v\rightarrow \Phi^{3}
\eeq 
we use deriving (3.14) in the preexponential factor as well as in quadratic 
terms in the exponent should be considered as an approximation. Terms with 
derivatives of fields could be added to r.h.s. of (3.32). However, possible 
corrections, for example,
\beq      
   \rho\nabla^{2}\Phi^{a}
\eeq 
are down by extra factor of g because $\qrho \sim\qg/\Phi^{2}$, see (2.11). In 
other words corrections (3.33) produce expansion in powers $\qrho/x^{2}\ll 1$. 
However, there are also other corrections to (3.32) like
\beq      
  \zeta_{\al} W^{\al a},
\eeq 
which are compatible with N=1 SUSY. These  are not suppressed by powers of 
coupling constant. These corrections could be recovered using N=2 SUSY. We will
 do it in the next section.

\section{Effective vertex and N=2 SUSY}
\setcounter{equation}{0}

In the last section we derived the effective instanton-induced vertex (3.14) 
making use of the classical expressions for superinstanton fields of section 2. 
However, certain terms in (3.14) (preexponential factor as well as terms 
quadratic in quantum fluctuations in the exponent) are beyond our control in the 
leading order of the
 semiclassical approximation. We fixed them using N=1 supersymmetric promotion 
rule (3.32).

Now we are going to check if our vertex (3.14) is N=2 supersymmetric. Of course, 
classical terms should be. But what about ``quantum pieces''? We will show that 
the rule (3.32) should be modified to be compatible with N=2 SUSY. This 
modification fixes possibles corrections of (3.34) type which remains
 undetermined in the previous section.

To made a check on N=2 SUSY of (3.14) we use N=2 superfield formalism. The N=2 
chiral superfield is a subject to certain constrains \cite{GSW}. Its low
 components can be written in the form
\beq      
 \bPsi^{a}(x,\tht_{1},\tht_{2})=\Phi^{a}(x,\tht_{1})+
   \sqrt{2}\tht_{2\al}W^{a\al}(x,\tht_{1})+\cdots ,  
\eeq
where $x$ is assumed to be a left-handed coordinate and $\tht_{1}$, $\tht_{2}$ 
are two $\tht$-parameters associated with the first and the second supersymmetry 
respectively. The anti-chiral field has the form
\beq      
 \bPsi^{a}(x^{R}_{N=2},\bth_{1},\bth_{2})=\bPh^{a}(x^{R}_{N=2},\bth_{1})+
   \sqrt{2}\bth^{\dal}_{2}\bar{W}^{a}_{\dal}(x^{R}_{N=2},\bth_{1})+\cdots ,  
\eeq 
where N=2 rihgt-handed coordinate is related to $x$ in (4.1) as 
\beq      
(x^{R}_{N=2})_{\mu}=x_{\mu}-2i\bth_{1}\bsi_{\mu}\tht_{1}-
     2i\bth_{2}\bsi_{\mu}\tht_{2},
\eeq 
while $\bth_{1}$, $\bth_{2}$ are conjugate $\tht$-parameters.

Let us check that exponent in (3.14) can be rewritten in terms of superfields 
(4.1), (4.2), at least classically. The identification (3.16), (3.17) (as well 
as transformation laws (2.14), (2.22)) shows that $-\al$ and $-\bet_{1}$ plays 
the role of N=1 SUSY $\tht$-parameters $\tht_{1}$ and $\bth_{1}$.The symmetry 
between $\la$ and $\psi$ suggests that parameter $\zeta$ should be identified 
with $\tht_{2}$. What about $\bth_{2}$? The symmetry between $\bbt$ and 
$\bar{w}$ (see (2.12), (2.13)) suggests that if we introduce a new 
intergration 
variable 
$\bet_{2}$ in (3.14) instead of $\bbt_{inv}$ as 
\beq      
\bbt_{inv\dal}=\frac{v}{2\sqrt{2}}(\bar{u}_{inv}\tau_{3}u_{inv}\bet_{2})_{\dal}
\eeq
(this should be compared with (2.17)) we can try to identify it with $\bth_{2}$.

It is easy to see that the first, the second, the forth and the last terms in 
the exponent in (3.14) can be combined into the simple expression
\beq      
-\frac{4\qpi}{\qg}\qrho_{inv}\bPsi^{a}\Psi^{a},
\eeq
where
\beq      
   \Psi^{a}=\Psi^{a}(x_{0},\tht_{1}=-\al, \tht_{2}=\zeta)
\eeq
and
\beq      
   \bPsi^{a}=\bPsi^{a}(x_{0\mu}-2i\bet_{1}\bsi_{\mu}\al-2i
   \bet_{2}\bsi_{\mu}\zeta,\bth_{1}=-\bet_{1}, \bth_{2}=\bet_{2})
\eeq

Let us explain (4.5) in more detail. The first term in the exponent in (3.14) is 
reproduced when we combine fields $\Phi$ and $\bPh$ in expansions (4.1) and 
(4.2) in (4.5). The second term comes from the W component of $\Psi$ times the 
$\bPh$ component of $\bPsi$ in (4.5). The forth term corresponds to the 
$\bar{W}$ component of the $\bPsi$ times the $\Phi$ component of $\Psi$. Here we 
make use of definition (4.4) and drop out the non-abelian commutator term 
proportional to $\ep^{abc}\bPsi^{a}\Psi^{b}$ which is unessential for our future 
purposes. Striktly speaking we get $\bar{W}^{3}v$ instead of 
$\bar{W}^{a}\Phi^{a}$ in (4.5) from the forth term in the exponent in (3.14). To 
get actually  $\bar{W}^{a}\Phi^{a}$ we use the promotion rule (3.32) as well as 
the gauge invariance.
 The last term in the exponent in (3.14) corresponds to the expansion of $\bPsi$ 
in (4.5) in powers of the shift
 $2i\bet_{2}\bsi_{\mu}\zeta$ of the N=2 right-handed coordinate, see (4.7).
 Note, that $\bar{W}W$ term in (4.5) cannot be checked at the classical
 level as it is quadratic in quantum fluctuations.

In the similar way the third and the fifth terms in (3.14) can be combined into 
expression 
\beq      
   -\frac{\sqrt{2}\qpi}{\qg}i\qrho_{inv}(\bar{\nabla}_{2}\bar{u}_{inv}\tau^{a}
    u_{inv}\bar{\nabla}_{1})\bPsi^{a},
\eeq
where
\beq      
  \bar{\nabla}_{2\dal}=\frac{\prt}{\prt \bth_{2}^{\dal}}
\eeq
is the covariant derivative with respect to the second supersymmetry, while 
$\bar{\nabla}_{1}$ is given by (3.15).

Combining (4.5) and (4.6) we arrive finally to the N=2 supersymmetric vertex
$$
V^{N=2}_{I}=-\frac{1}{4\pi^2}\Lambda^4\int d^4 x
\frac{d\rho}{\rho}\frac{d^3
u_{inv}}{2\pi^2}\frac{1}{\Psi^{a4}}d^{2}\alpha d^2\zeta d^2\bar{\eta_{1}}
d^{2}\bar{\eta_{2}}
$$
\beq
exp\left[-\frac{4\pi^2}{g^2}\rho^{2}_{inv}\bar{\Psi}^{a}\Psi^{a}
-
\frac{\pi^{2}}{\sqrt{2}g^2}\rho^{2}_{inv}
i(\bar{\nabla}_{i}\bar{u}_{inv}\tau^{a}u_{inv}\bar{\nabla}^{i})\bar{\Psi^{a}},
\right]
\eeq

where i=1,2 denotes the first and the second supersymmetry and
$$
(\bar{\nabla}_{i}\bar{u}_{inv}\tau^{a}u_{inv}\bar{\nabla}^{i})=
  2(\bar{\nabla}_{2}\bar{u}_{inv}\tau^{a}
    u_{inv}\bar{\nabla}_{1}).
$$
The arguments of superfields here are given by (4.6), (4.7). To get the factor 
$1/\Psi^{4}$ in the exponent in (4.10) we change our promotion rule (3.32) to 
\beq
v\rightarrow\Psi^{3},
\eeq
which is obviously compatible with N=2 SUSY. The SUSY invariant size 
$\rho_{inv}$ should be understood in (4.10) as follows (see change of variables 
(4.4))
\beq
  \qrho_{inv}=\qrho[1+\sqrt{2}i\Psi^{a}(\bet_{1}\bar{u}_{inv}\tau^{a}
    u_{inv}\bet_{2})]
\eeq
Here we use (4.11) once again.

Eq. (4.10) is our final result for the instanton-induced effective vertex in N=2 
SUSY gauge theory. Note, that we derived (4.10) from (3.14) using only low 
components of superfields $\Psi$, $\bPsi$ (see (4.1), (4.2)). However, the 
expression (4.10) is more general. It contains also higher components which 
cannot be restored in the large distance limit.The reason is that they contain 
terms like $\nabla^{2}\Phi$ which give $\dl$-functional contributions to the 
correlation function (3.19) and therefore are unvisible in the large distance 
limit.

In the conclusion of this section let us summarize approximations we use to 
obtain (4.10). First, we assumed the weak coupling regime and pick up the 
leading effect in $\qg$. Second, we assumed the large distance limit
 $x^{2}\gg\qrho$ because the way we derived the effective vertex assumes 
checking the large distance behaviour of instanton field (see section 
3)\footnote{This condition probably could be relaxed, see the discussion on page 
12. We come back to this point in the conclusion.}. Also our promotion rule 
(4.11) could contain corrections (3.33) which gives powers of $\qrho /x^{2}$.

Besides this, for the sake of simplicity we do not include in (4.10) certain 
terms related to the non-abelian structure of our theory. These effects plays no 
role in the low-energy effective Lagrangian we are aiming to derive using (4.10) 
in the next section. Namely, we do not write explicitly gauge factors 
$e^{-2V_{g}}$ in (3.14) and (4.10) as well. These factors can be easily restored 
using gauge invariance. We also drop out the term proportional to 
$\ep^{abc}\bPsi^{a}\Psi^{b}$ in the exponent in (4.10). This term can be 
extracted from (3.14) in a strightforward manner.

Finally, we assumed condition $x^{2}\ll v^{2}$, in order not to deal with 
exponential tails of instanton field. This condition can be easily relaxed. In 
the next section we go to the limit $x^{2}\gg v^{2}$ and derive low-energy 
vertex for light fields using (4.10).

Let us note, that our result in eq. (4.10) could be obtained in a much more 
simple way by N=2 supersymmetrization of the effective vertex (3.8) for the 
gauge-Higgs theory. To see this observe, that term (4.8) is the N=2 
supersymmetrization of the Callan-Dashen-Gross effective vertex, while term 
(4.5) is a direct N=2 supersymmetrization of the first term in the exponent 
(3.8).
  
\section{Low energy effective Lagrangian}
\setcounter{equation}{0}

In this section we are going to derive the low energy effective Lagrangian 
starting with our effective vertex (4.10) for microscopic theory. In particular, 
we recover Seiberg-Witten result (1.1) in which the one-instanton contribution 
to the prepotential is
\beq
 {\cal F}_{I}(A)=-\frac{1}{8\qpi}\frac{\La^{4}}{A^{2}},
\eeq
as a leading order effect in the large distance expancion.

So far we considered large but not very large distance limit 
  $\qrho\ll x^{2}\ll 1/v^{2}\sim 1/M^{2}_{W}$. Now we are going to relax the 
condition $x^{2}\ll 1/v^{2}$. It is clear that to do so we have to add the 
instanton-induced vertex (4.10) to the tree-level Lagrangian (2.3) and consider 
gauge symmetry breaking in this theory expanding field $\Psi^{a}$ around its vev
\beq
  \Psi^{a}=\dl^{a3}v + \dl\Psi^{a}
\eeq
Then $\Psi_{1}$  and $\Psi_{2}$ components become massive and do not propagate
 at large distances. To write down the low energy effective Lagrangian for light 
fields in $\Psi_{3}$ supermultiplet at distances  $x^{2}\gg 1/v^{2}$ we have to 
integrate out heavy fields in (2.3) with (4.10) added.

For one-instanton induced effects this boils down to dropping out heavy fields 
in (4.10). We have for the low energy effective vertex
$$
V^{LE}_{I}=-\frac{1}{4\pi^2}\Lambda^4\int d^4 x
\frac{d\rho}{\rho}\frac{d^3
u_{inv}}{2\pi^2}\frac{1}{\Psi^{4}_{3}}d^{2}\tht_{1} d^2\bth_{1} d^2\tht_{2}
d^{2}\bth_{2}
$$
\beq
exp\left[-\frac{4\pi^2}{g^2}\rho^{2}_{inv}\bar{\Psi}_{3}\Psi_{3}
-
\frac{\pi^{2}}{\sqrt{2}g^2}\rho^{2}_{inv}
i(\bar{\nabla}_{i}\bar{u}_{inv}\tau^{3}u_{inv}\bar{\nabla^{i}})\bar{\Psi_{3}}
\right]
\eeq

Grassmann parameters here are redefined to their conventional notations using 
(4.6), (4.7), while
$$
 \rho_{inv}^2=\rho^2 [1-
 \sqrt{2}i\Psi_{3}(\bar{\tht_{1}}\bar{u}_{inv}\tau_{3}u_{inv}\bar{\tht_{2}})]
$$
\beq
= \rho^2 [1+
\frac{1}{\sqrt{2}}
i\Psi_{3}(\bar{\tht_{i}}\bar{u}_{inv}\tau_{3}u_{inv}\bar{\tht^{i}})]
\eeq

Corrections to (5.3) involve loop graphs with vertex (4.10) inserted as well as 
perturbative interactions from (2.3). In these graphs light fields are in
 external legs 
while heavy fields propagate in loops. As we explained in section 3 these graphs 
will give corrections $(\qrho/x^{2})^{n}$ to the leading large distance 
behaviour of correlation functions. These effects can be encoded in (5.3) as an 
insertions of powers of the operator (see(3.33))
\beq
  \rho\nabla^{2}\sim g\frac{\nabla^{2}}{\Psi_{3}}
\eeq
 We ignore these effects as they are suppressed in coupling 
constant. Thus, to the leading order in $g^{2}$ we are left with the low energy
 vertex (5.3).

Let us now note, that our result in (5.3) make a bridge between two approaches 
to instanton calculations in SUSY theories. The first one developed in 
\cite{NSVZ, FS} use N=1 superfield formalism to calculate certain correlation 
functions
in the instanton background. Instanton superfields in this approach are written 
down in manifestly supersymmetric way in terms of SUSY invariant
 combinations. We follow this method in section 2.
 In refs. \cite {NSVZ, FS} certain correlation functions of chiral superfields 
were calculated. However, the naive extrapolation of this method to correlation 
functions of anti-chiral fields would give wrong results.

Now we know the reason for this. Besides classical effects our effective vertex 
(5.3) (or (4.10)) contains certain ``quantum'' effects which appear when we use 
the promotion rule (4.11). In particular, the promotion of $1/v^{4}$ to 
$1/\Psi^{4}$ in the preexponent of (4.10), (5.3) has no effects on expressions 
for chiral fields. However, in correlation functions of anti-chiral fields this 
give an additional ``quantum'' contributions. Surprisingly, they appear to be of 
the same order as classical ones after intergrating over instanton size. The 
reason is that classical expressions of fields are proportional to $\qrho$, 
however typical $\qrho \sim \qg /v^{2}$. Thus, they are of the same order. The 
effective vertex (4.10), (5.3) encorporates both classical and ``quantum'' 
effects in manifestly supersymmetric way.

Another approach originated by Affleck-Dine-Seiberg \cite{ADS} (see also 
\cite{FP, DKM}) deals with instanton fields in components. In particular, in 
order to extract the Seiberg - Witten term (5.1) the correlation function of
 anti-chiral fermions $<\bla\bla\bar{\psi}\bar{\psi}>$ has been considered 
\cite{S, FP}.
 Although components of chiral fields can be arranged in supermultiplets
 (2.27),
 (2.28) components of anti-chiral fields in this approach differ
 from the expansion of our formulas (2.40), (2.41) in components.
 The difference is that they are proportional to $\qrho$, rather then 
$\qrho_{inv}$ as in our approach. Thus, the supersymmetry is spoiled. In fact, 
it should be restored after the integration over collective coordinates.

We can mimic this approach in our effective Lagrangian method writing down the 
vertex
$$
V'=-\frac{\Lambda^4}{4\pi^2}\int d^4 x
\frac{d\rho}{\rho}\frac{d^3
u}{2\pi^2}d^{2}\tht_{1} d^2\bth_{1} d^2\tht_{2}
d^{2}\bth_{2}  \frac{1}{v^{4}}
$$
\beq
exp-\frac{4\pi^2}{g^2}\rho^{2}\left[(1-\sqrt{2}iv\bth_{1}\bar{u}\tau_{3}
 u\bth_{2})\bar{\Psi}_{3}v + \bPsi_{3}\dl\Psi_{3}
\right]
\eeq
where $\Psi_{3}$ is expanded as in (5.2).

The vertex (5.6) differs from our vertex in (5.3) in two ways. First, 
$\bPsi_{3}\dl\Psi_{3}$ comes multiplied by $\qrho$ in (5.6), hence the 
expressions for anti-chiral fields produced by (5.6) are proportional to 
$\qrho$, rather then $\qrho_{inv}$ as we discussed above. Second, no promotion 
rule is used in the preexponent, as well as in definition of  $\qrho_{inv}$ in 
front of $\bPsi_{3} v$. Thus (5.6) contains only classical effects.

Now we show that (5.6) reduces to (5.3) at least at the Seiberg - Witten order 
(the leading order in derivatives of fields). To do this rewrite the exponent in 
(5.6) as
\beq 
exp-\frac{4\pi^2}{g^2}\rho^{2}\left[\bPsi_{3}\Psi_{3} -
\sqrt{2}i(\bth_{1}\bar{u}\tau_{3}
 u\bth_{2})\bar{\Psi}_{3} v^2
\right]
\eeq
Then make a change of variables
\beq
\bth'_{2} = \bth_{2}\frac{v^2}{\Psi_{3}^{2}}.
\eeq
We get 
$$
V'=-\frac{\Lambda^4}{4\pi^2}\int d^4 x
\frac{d\rho}{\rho}\frac{d^3
u}{2\pi^2}d^{2}\tht_{1} d^2\bth_{1} d^2\tht_{2}
d^{2}\bth_{2}'
$$
\beq
\frac{1}{\Psi^{4}_{3}}exp\left[ -\frac{4\pi^2}{g^2}\rho_{inv}
 \bPsi_{3}\Psi_{3}
\right],
\eeq
with $\qrho_{inv}$ from (5.4). This is nothing else then our vertex (5.3) with 
Callan - Dashen - Gross term omited (it contains higher derivatives of fields). 
In fact, the discussion above can be viewed as a purely ``classical derivation'' 
of our promotion rule (4.11). Note, that the change of variables (5.8) is 
possible only in the leading order of large distance limit when the dependence 
of $\bPsi_{3}$ on its argument $\bth_{2}$ is ignored.

Thus we have shown that our effective vertex (4.10), (5.3) combine the 
superfield formalism of refs. \cite{NSVZ, FS} with the component approach in 
refs. \cite {ADS}.

Now let us integrate over the size instanton $\rho$ in (5.3). The dependence on 
$\bth_{1}$, $\bth_{2}$ in (5.3) comes from explicit dependence of $\qrho_{inv}$ 
on these parameters as well as the dependence of $\bPsi_{3}$ on its arguments. 
According to this, there are three ways to saturate the integrals over  
$\bth_{1}$, $\bth_{2}$.

First, ignore the $\bth$-dependence of $\bPsi$. This will give us the leading
 Seiberg - Witten term. Once the dependence of  $\bth_{1}$, $\bth_{2}$ comes 
only via  $\qrho_{inv}$ we can write down
\beq
 \int \qd\bth_{1}\qd\bth_{2}f(\qrho_{inv})= -\frac{1}{2}\Psi^{2}_{3}
 \left[(\qrho\frac{\prt}{\prt\qrho})^{2}
 - \qrho\frac{\prt}{\prt\qrho}\right]f(\qrho)
\eeq
for any function f.

We see that the integral over $\rho$ reduces to a total derivative. Moreover,
 infinite size instanton do not contribute in (5.3) and we are left with zero 
size instanton. This important phenomenon was first noticed in \cite{NSVZ}. 
Substituting (5.10) into (5.3) we have
\beq
V^{LE}_{F}=\frac{\Lambda^4}{16\pi^2}\int d^4 x
d^{2}\tht_{1} d^2\tht_{2}\frac{1}{\Psi^{2}_{3}}.
\eeq
Note that (5.11) is N=2 F-term.

This is nothing else then Seiberg - Witten one-instanton
vertex. To see this rewrite (5.11) in 
N=1 supermultiplets. It reads
\beq
V^{LE}_{F}=\frac{\Lambda^4}{8\pi^2}\int d^4 x
d^{2}\tht_{1} d^2\bth_{1}\frac{\bar{A} A}{A^{4}} -
\frac{3 \Lambda^4}{16\pi^2}\int d^4 x
d^{2}\tht_{1}\frac{W^{2}_{3}}{A^{4}}.
\eeq
where $A\equiv \Phi_{3}$. This coincides with (1.1) where function
 $\cal F_{I}$ is given by (5.1).

Now let us work out other contributions in (5.3). The second possibility
 is to extract one power of $\bth_{1}\bth_{2}$ from $\bPsi_{3}$ and another 
power from $\qrho_{inv}$. It is easy to see that integral over $\rho$ again 
reduces to the total derivative. However, in this case the function $f$ is zero 
in both limits $\rho=0$ and $\rho=\infty$.

The last possibility is to ignore $\bth$-dependence in $\qrho_{inv}$ and 
saturate integrals using  $\bth$-dependence of $\bPsi_{3}$ in (5.3). Then the 
integral over $\rho$ gives
$$
V^{LE}_{D}=\frac{\Lambda^4}{8\pi^2}\int d^4 x
d^{2}\tht_{1} \qd\bth_{1} \qd\tht_{2} d^2\bth_{2}\frac{\td u}{2\qpi}
$$
\beq
\frac{1}{\Psi^{4}_{3}}\log [\bPsi_{3}\Psi_{3}+
\frac{i}{4\sqrt{2}}(\bar{\nabla}_{i}\bar{u}\tau_{3} u\bar{\nabla}^{i})\bPsi_{3}]
\eeq
This is N=2 D-term in contrast to the leading Seiberg - Witten effect (5.11), 
which is F-term. The factor $1/\Psi^{4}_{3}$ in front of logarithm account 
correctly for the anomalous chiral symmetry violation induced by instanton. 
Note, that the dependence on the orientation vector $u_{\mu}$ enters only in the 
Callan - Dashen - Gross piece.

Eqs. (5.11) and (5.13) is our final result for the one-instanton low-energy 
effective Lagrangian in the theory at hand. Anti-instanton generate the complex 
conjugate to (5.11), (5.13).

The D-term (5.13) is the correction in derivatives of fields to the leading 
Seiberg - Witten F-term (5.11). In general the next-to-leading correction
to the Seiberg-Witten effective theory (1.1) is given by N=2 D-term 
\cite{Henn}
\beq
S^{LE}_{next-to-leading}=\int d^4 x
d^{2}\tht_{1} \qd\bth_{1} \qd\tht_{2} d^2\bth_{2} K(\Psi ,\bar{\Psi}),
\eeq
where $K$ is a real function of its arguments.  In N=1 superfields (5.14)
reads
$$
S^{LE}_{next-to-leading}=\frac{1}{16}\int \fd x\qd\tht_{1}\qd\bth_{1}
\left(
K_{A\bar{A}}[\qnbl A\qbnb\bar{A}
+ 2\bnbl^{\dal}\nbl_{\al} A\nbl^{\al}\bnbl_{\dal} A \right.
$$
$$
+4\nbl_{\al} W^{\al}_{3}\bnbl^{\dal}\bar{W}_{3\dal} - 
4\nbl_{(\al} W_{3\bt )}\nbl^{(\al} W^{\bt )}_{3} -
4\bnbl^{(\dal}\bar{W}^{\dbt )}_{3}\bnbl_{(\dal}\bar{W}_{3\dbt )}-
$$
$$
2\qnbl W^{2}_{3}-2\qbnb \bar{W}^{2}_{3}] -2K_{AA\bar{A}}W^{2}_{3}\qnbl A
-2K_{A\bar{A}\bar{A}}\bar{W}_{3}^{2}\qbnb\bar{A}+
$$
\beq
\left.
K_{AA\bar{A}\bar{A}}[-8W_{3\al}\nbl^{\al} A\bar{W}^{\dal}_{3}
\bnbl_{\dal}\bar{A}+4W^{2}_{3}\bar{W}^{2}_{3}]
\right).
\eeq

The subscripts on K here  denotes derivatives with respect to 
its arguments.

Comparing (5.15) with   (1.1) we see that the expansion goes in powers of 
operators
\beq
  \frac{\qnbl}{A},\;\;  \frac{W^{2}_{3}}{A^{3}}.
\eeq
If we ignore for a moment the Callan-Dashen-Gross term in (5.13) our result
 implies that the one-instanton contribution to
 the function $K(A,\bar{A})$ is given by
\beq
K_{I}(A,\bar{A})=\frac{\La^{4}}{8\qpi}\frac{1}{A^{4}}\log \bar{A} A + c.c.
\eeq

The Callan - Dashen - Gross term in (5.13) represents further corrections to
 (5.15) in the same parameters (5.16).

To conclude let us sum up our approximations. Our effective Lagrangian in (5.13) 
controls all powers of corrections in derivatives (5.16) to the leading Seiberg 
- Witten term (5.11).
Integrating over orientations in (5.13) will produce the infinite series in 
powers of this parameters. Instead, the corrections of type (5.5) which involved 
extra powers of coupling constant are beyond our control. The later effects 
becomes important at distances $x^{2}\sim \qrho$, which is much less then
the limit $x^{2}\gg\ 1/v^{2}\sim 1/M^{2}_{W}$
 assumed for a low energy effective theory to describe the physics of light 
degrees of freedom correctly. 

\section{Conclusions}
\setcounter{equation}{0}

The  Seiberg -Witten exact solution for the 
low energy effective Lagrangian of the
N=2 SUSY gauge theory 
 accounts for all possible powers of coupling constant $\qg$ and instanton 
factor $\La^{4} / A^{4}$ in the leading order of the
 large distance expansion.
	Instead in this paper we  limit ourselves in the weak coupling region
of the moduli space
 taking into account only one-instanton-induced effects in the leading order in 
$\qg$ and study the large
 distance expansion of low-energy theory. Our effective Lagrangian (5.13) 
contains all orders of corrections in derivatives of fields to the Seiberg - 
Witten term (5.11). We ignore only corrections in powers of
 $\qrho / x^{2}$, which are really corrections (5.5) in 
$\qg$ , provided large distance limit $x^{2}\gg 1/v^{2}$ is considered.

In particularly, the next-to-leading order correction to the Seiberg-Witten
exact solution
in the large distance
expansion is given by eq.(5.14), (5.15) \cite{Henn}. We have calculated here
the one instanton contribution to the function $K(A,\bar{A})$ (5.17).
The general form of the perturbative
contribution to this function has been studied in recent paper \cite{WGR}.
  For the abelian
low energy theory at hand we have
\beq
K(A,\bar{A})=c_{00}\log\bar{A}\log A,
\eeq
where $c_{00}$ is unknown constant.
Comparing (6.1) and (5.17) one might expect that in general function 
 $K(A,\bar{A})$ is given by the expansion
\beq
K(A,\bar{A})=\sum_{n,m=0}^{\infty}\bar{L}_{n}c_{nm}L_{m},
\eeq
modulo the real part of the holomorphic function which does not
contribute to (5.15). Here coulomb vector $L_{n}$ is defined as follows
\beq
L_{0}=\log A ,\;\; L_{n}=\frac{\Lambda^{4n}}{A^{4n}}, n=1, 2\ldots
\eeq
and coefficients $c_{nm}$ satisfy the reality condition $\bar{c}_{nm}=
c_{mn}$.
The expansion in (6.2) is invariant under $Z_{8}$ group (up to
Kahler transformation). Each $L_{n}$ 
with $n\ge 1$ comes from the instanton with the topolpgical carge $n$,
while $\bar{L}$'s come from anti-instantons. In particular, our result
(5.17) gives $c_{01}=1/8\pi^2$.
 In principle, 
 coefficients $c_{nm}$ could receive further logarithmic
corrections due to higher loops.

It would be  interesting to study the behaviour of function 
 $K(A,\bar{A})$ in
 the strong coupling region of the modular space.
Of particular interest is to find the singularities of this function.
 The reason is that the low energy theory is not
well defined in these singular points due to the absence 
of the  proper kinetic term. The duality transformation transforms (5.15)
into the same expression with new function $K_{D}$ defined as \cite{Henn}
\beq
K_{D}(A_{D},\bar{A}_{D})=K(A,\bar{A}),
\eeq
where dual field is $A_{D}={\cal F}'(A)$. Particularly dangerous effect
could occur if $K_{D}$ has a singularity at $a_{D}=0$ or $a_{D}=a$
where monopole or dyon
becomes massless. This would spoil the nice physical description \cite{SW}
of the low energy theory  as a dual massless
QED near these points.

Let us now say a few words about  $\qrho / x^{2}$ corrections to (5.3).
In fact, the natural conjecture is that our effective vertex (4.10) for the 
microscopic theory provides the systematic method to calculate all terms in the 
long-distance expansion of low-energy effective Lagrangian, including
 $\qrho / x^{2}$ corrections. The reason for that conjecture is that, as we 
explained in section 3 \cite{Z}, $\qrho / x^{2}$ corrections to the leading
at large distances
 behaviour of instanton field appears in instanton-induced effective
 Lagrangian method as loop diagrams. Still one may doubt in the above 
conjecture because actually we derived our effective vertex (4.10) using 
not only 
classical considerations but  the promotion rule (4.11) as well 
which could provide additional $\qrho / x^{2}$ corrections. However, we showed 
in the last section that the promotion rule (4.11) does not account for true 
quantum corrections. It can be understood as the result of a change of variables 
within the purely classical analysis.

In particularly, we can argue now that
 the Seiberg - Witten N=2 F-term (5.11) does not get any $\qrho / x^{2}$ 
corrections at all. To see this recall that F-term appears when we integrate in 
(5.3) over $\bth_{1}$ and $\bth_{2}$ using the explicit dependence of 
$\qrho_{inv}$ on this parameters. This insures that the integral over $\rho$ 
reduces to the integral of total derivative. Hence, the result for F-term comes 
from zero size instanton $\qrho \rightarrow 0$, so the corrections in $\qrho / 
x^{2}$ are zero.

The author is grateful to N. Dorey, V. Khoze
 and M. Mattis for valuable discussions and a particularly close contact
 at the early stage of this work, to 
 Theoretical Physics group at the University of Swansea for
 hospitality and to the Higher Education Funding Council for Wales for support.
This work was also supported by the Russian Foundation for 
Fundamental Studies under Grant No. 96-02-18030.

    \end{document}